\documentclass[a4paper]{article}
\usepackage[margin=25mm]{geometry}
\usepackage{amsmath}
\usepackage{amsfonts}
\usepackage{amssymb}
\usepackage{graphicx}
\usepackage[utf8]{inputenc}
\usepackage{authblk}
\usepackage{imakeidx}
\usepackage{tabularx}
\usepackage[sorting=none]{biblatex}

\DeclareUnicodeCharacter{0301}{\'{e}}
\DeclareUnicodeCharacter{0413}{\'{e}}
\DeclareUnicodeCharacter{043E}{\'{e}}
\DeclareUnicodeCharacter{043B}{\'{e}}
\DeclareUnicodeCharacter{043F}{\'{e}}
\DeclareUnicodeCharacter{0440}{\'{e}}
\DeclareUnicodeCharacter{0438}{\'{e}}
\DeclareUnicodeCharacter{0441}{\'{e}}
\DeclareUnicodeCharacter{0442}{\'{e}}
\DeclareUnicodeCharacter{0430}{\'{e}}
\DeclareUnicodeCharacter{043D}{\'{e}}
\DeclareUnicodeCharacter{044C}{\'{e}}
\DeclareUnicodeCharacter{043A}{\'{e}}
\DeclareUnicodeCharacter{0439}{\'{e}}
\DeclareUnicodeCharacter{043C}{\'{e}}
\DeclareUnicodeCharacter{0456}{\'{e}}
\DeclareUnicodeCharacter{0446}{\'{e}}
\DeclareUnicodeCharacter{0435}{\'{e}}
\DeclareUnicodeCharacter{0437}{\'{e}}
\DeclareUnicodeCharacter{044F}{\'{e}}

\providecommand{\keywords}[1]
{
  \small	
  \textbf{\textit{Keywords---}} #1
}

\title{Innovative photodetector for LIDAR systems}

\author[1,2]{K. Khuseynzada}
\author[2,3]{A. Sadigov}
\author[2]{J. Naghiyev}
\author[2,3]{F. Ahmadov}
\author[2,3]{G. Ahmadov}
\affil[1]{Mingechevir State University}
\affil[2]{National Nuclear Research Center}
\affil[3]{National Academy of Sciences}

\date{August 2022}

\begin{document}

\maketitle

\begin{abstract}
Optical imaging systems are widely used in the drone industry (optical navigation, terrain scanners), industry (distance measurement, 3D visualization), security (optical scanners and motion sensors), robotics (motion sensors, terrain or field scanning), etc. Lidar (Light Detection and Ranging) systems have a different structure depending on their purpose, and the principle of operation is based on measuring the time difference between the emission time of a laser or other light source and the time of its reflection from the object, where the reflected photons fall on the photodetector. This time interval is about hundreds of ps. Accuracy, photography speed and system efficiency depend on the performance of the photodetector module. Modern development of high technologies; the new efficiency of detecting a large photon flux makes it possible to develop photodetectors with a nanolayer micropixel structure. The paper presents the development of a modern and highly sensitive micropixel avalanche photodetector. The developed photodetector have high speed, low noise and high resolution. The improvement of these parameters allows the developed photodetectors to become an indispensable component of lidar systems. 
\end{abstract}


\section{Introduction}
LIDAR is the most modern and efficient technology for obtaining more accurate and complete digital information about the terrain, creating digital models of the earth's surface relief. As a result of laser scanning, digital relief models (DRM) are created with higher accuracy with the necessary information, and this technology is more efficient than photogrammetric methods. In LIDAR technology, the determination of two parameters is of particular importance: flight time and luminous flux intensity. Flight time is the time between the moment that the laser emits light and the moment the laser beam arrives back from the target. This time is calculated in the following form: \(t = 2 * d / c\), where d is the distance to the target and c is the speed of light. The distance to the target is calculated by the formula d = t * c / 2. The intensity of the light flux depends on the size and speed of the target. The intensity of the light flux returning from the target depends on the amplitude recorded by the photo detector. It is these parameters that make it possible to obtain accurate information about the purpose and widespread use of such systems. For example, with the help of LIDAR technology, all visible objects, including building facades, bridges, road, streets, flora, etc., can be seen from a moving vehicle in a short time along a wide corridor. On the other hand, LIDAR can provide environmental, land-use and mapping services as well as wide-corridor planning and high-precision DRM generation using aircraft systems. LIDAR systems provide all the terrain information you need, covering all objects, elements and vegetation on the earth's surface, and can scan the earth's surface beneath an aircraft with high accuracy. Depending on the system, the scan rate can vary from 50,000 to 100,000 pulses per second, creating a dense cloud of 3 criterion points. Additional data, such as laser intensity values, can be used to determine the type of surface, such as soil, vegetation, and buildings. Although the system is widely used in various fields, the improvement of LIDAR devices is an urgent problem for many research institutes, and research mainly aimed at improving the photodetector part of the system, actualy the photodetector module. Until recently, vacuum photomultiplier tubes (PMTs) used as photodetectors. Semiconductor-based photomultipliers (SiPM), an alternative to conventional vacuum PMT, have already replaced PMTs in many areas. Thus, most modern optical recording systems are based on silicon photodetectors, for example, in high-energy physics research (CERN ATLAS, COMPASS, etc.)\cite{akbarov2018scintillation}, \cite{akbarov2019fast}, \cite{holik2020program}, \cite{sadygov2012development} equipment for PET tomography in medicine\cite{ahmadov12new}, \cite{holik2020miniaturized}, nuclear spectrometers\cite{akbarov2020scintillation}, \cite{ahmadov2017new}, LIDAR systems etc\cite{sadigov2017micropixel}, \cite{sadigov2016new}. Leading manufacturers and research centers offer avalanche photodetectors of various designs depending on the application\cite{sadigov2012new}, \cite{ahmadov2013micro}, \cite{ahmadov2014micro}. These photodetectors (photodiodes) are widely used in two main types - surface pixels and internal pixel structures. Both types consist of individual micropixels of a passive element (microresistor or microcapasitors) placed on a common base. Each pixel of the photodiode operates in Geiger mode, and the photodiode has the only photoelectronic recording capability. Separate passive pixels used to disable the avalanche process. Parallel-connected micropixels ensure linearity of the photodetector. An Azerbaijani scientist Z.Sadygov first proposed SiPMs\cite{sadygov2020silicon}, and patented basic designs\cite{sadygov2016multi}. During the last 10 years, the different modifications of MAPD photodetectors were studied\cite{ahmadov2020new}, \cite{nuruyev2020performance}, \cite{sadigova2021improvement}, \cite{nuriyev2018performance}.  Independent p-n junction (micropixels) in the first structure. This is the type of structure known as a silicon photomultiplier or micropixel photo counter (MPPC) and is commercially available. One drawback is the small linear amplitude range associated with low pixel density due to the pixels and microresistors placed on the surface of the photodiode. Therefore, an increase in the pixel density above 1000 pixels / $mm^2$ leads to a significant decrease in the sensitive area of the photodetector. This, in turn, leads to deterioration in the efficiency of photon detection and amplitude resolution. Therefore, at present, one of the main directions for researchers is to increase the efficiency of photon detection, resolution, linearity of speed (recovery time) and amplitude of avalanche photodiodes.

\section{Literature review}

Devices for detecting and processing optical information are used in many scientific and household devices. The key element of such devices is a photodetector, which converts optical information into an electrical signal. The main operating parameters of a photodetector, such as sensitivity and speed, determine the efficiency of such devices. Traditionally, vacuum photomultipliers are used in such optical devices. However, in recent years, semiconductor photomultipliers have been developed that are adequate analogues of vacuum photomultipliers.

At present, semiconductor photomultipliers have become commercially available and can be used to detect single light quanta in the visible and IR regions of the optical spectrum. Semiconductor photomultipliers consist of many independent cells in which the Geiger regime of amplification of photoelectrons is carried out. As a result, a unique combination of fast photoresponse (photoresponse duration about 5 ns) and high signal amplification (about $10^6$) is achieved. However, to solve a number of applied problems, the rise time of the photoresponse is less than 1 ns. In addition, a large photosignal gain in semiconductor photomultipliers leads to an undesirable effect - optical crosstalk. This effect is due to the fact that a large gain (about $10^6$) of the photosignal is accompanied by the emission of optical photons in the avalanche region of the semiconductor. These photons are absorbed in neighboring cells of the photodetector and cause a false start of the avalanche process. Therefore, it is necessary to reduce the photosignal avalanche amplification factor to $10^4$, which is insufficient for operation in the single photoelectron counting mode.

Various modifications of highly efficient photodetectors are known and commercially available. For example, there is a photodetector that includes a semiconductor substrate, on the surface of which a matrix of semiconductor regions is made, forming a p-n junction with the semiconductor substrate. The surface of the semiconductor regions contains a resistive layer with a certain conductivity and a field electrode translucent to light. Avalanche amplification of photoelectrons is carried out at the boundary of the semiconductor substrate with semiconductor regions. In this case, the avalanche current flows to the translucent field electrode through the resistive layer located above these areas. The disadvantage of the device is the low quantum yield of the device in the visible region of the spectrum due to the low transparency of both the resistive layer and the semiconductor regions. In addition, the photoelectrons formed between the semiconductor regions do not have the opportunity to be amplified, which leads to a decrease in the photocurrent gain in the device.

The other device is known, including a semiconductor substrate of n-type conductivity, on the surface of which a resistive layer with a certain conductivity, a dielectric layer and a semiconductor epitaxial layer of p-type conductivity are sequentially located. Separate highly doped n-type semiconductor regions are formed inside the dielectric layer, having access on one side to the resistive layer, and on the opposite side to the epitaxial layer. Highly doped regions of n-type conductivity provide localization of the avalanche process in p-n junctions, separated from each other by regions of the dielectric layer. The photosensitive layer in which photoelectrons are created is an epitaxial layer grown on the surface of foreign materials - dielectric and resistive layers. Therefore, the main disadvantages of the device are the complexity of the technology for manufacturing such epitaxial layers and the high level of dark current, which leads to a deterioration in the sensitivity and signal-to-noise ratio of the device.
The other device is also known, including a semiconductor layer, on the surface of which there are many semiconductor regions that form potential barriers in the form of p-n junctions with a semiconductor layer. Individual microresistors connect the semiconductor regions to a common conductive bus separated from the semiconductor layer by a dielectric layer. In the device, each semiconductor region (pixel), independently of the others, can operate in a mode above the breakdown potential, that is, each pixel in the Geiger counter mode. Therefore, the photocurrent gain in the device may exceed $10^6$. However, as mentioned above, the use of the device at such high gains is difficult due to the appearance of optical crosstalk. This is the first major drawback of the prototype. The second main drawback of the device is insufficiently high performance due to the high capacitance of both the pixel itself and parasitic capacitances in the device. Here it should be noted that for a given pixel that amplifies the photocurrent, all other pixels that do not participate in the amplification of the photocurrent are parasitic capacitance.

\section{Concept of photodetector}
\subsection{Structure}

The proposed concept of a new photodetector is aimed at reducing the level of optical crosstalk and improving the performance of a semiconductor avalanche detector. To achieve these technical results in a semiconductor avalanche detector, including a semiconductor layer, on the surface of which a plurality of semiconductor regions are formed, forming potential barriers with a semiconductor layer, a common conductive bus separated from the semiconductor layer by a dielectric layer, and individual microresistors connecting the semiconductor regions with a common conductive bus, additionally formed new elements. These new elements are individual emitters formed on the surface of said semiconductor regions, an additional conductive bus, and additional individual microresistors connecting the individual emitters to the additional conductive bus. Said semiconductor layer is either used independently to create a photodetector or it is formed by epitaxial growth on the surface of semiconductor or dielectric substrates. Then, the necessary elements are formed on the surface of the semiconductor layer.

The concept is illustrated in figure 1, which shows a cross section of a semiconductor avalanche detector. The proposed semiconductor avalanche detector contains a semiconductor layer 1, on the surface of which a plurality (matrix) of semiconductor regions 2 is formed, forming potential barriers in the form of a p-n junction with a semiconductor layer. Each semiconductor region has an individual microresistor 3 connecting it to a common conductive bus 4. The microresistors and the conductive bus are isolated from the semiconductor layer 1 by a dielectric layer 5. Individual emitters 6 are formed on the surface of the mentioned semiconductor regions. Individual emitters are connected to an additional conductive bus 7 by means of additional individual microresistors 8. The device has a contact 9 to the semiconductor layer.

\begin{figure}[h]
    \centering
    \includegraphics[width=0.6\textwidth]{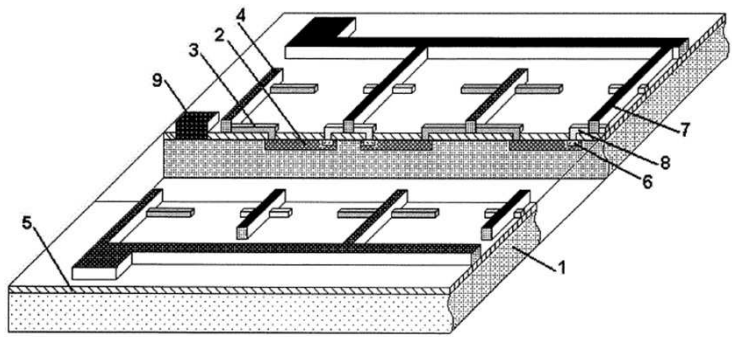}
    \caption{Schematic view of the photodetector concept}
    \label{fig1}
\end{figure}

\subsection{Working principal}

The device works as follows. A potential with a polarity corresponding to the reverse bias of the p-n junction formed between the semiconductor layer 1 and the semiconductor regions 2 is applied to the semiconductor layer 1 relative to the conductive tires. Due to the small size (about 50 u × 50 u) of the semiconductor regions, charge carriers are not always present in the depletion layer, and therefore such small area p-n junctions (or pixels) can operate in a mode above the breakdown potential by 2-5 V. In the absence of a photoelectron (or dark charge carriers), the pixel potential is equal to the individual emitter potential, and therefore the current through the emitter is zero. In the case of the appearance of a single photoelectron in the depleted region of the pixel, an avalanche process occurs and the excess voltage, that is, $\vec{\Delta}$ V about 2-5 V, drops in an individual microresistor. In this case, the potential $\vec{\Delta}$V about 2-5 V completely opens the potential barrier between the pixel (semiconductor region) and the individual emitter, as a result, an enhanced current flows through the individual emitter, which can only be limited by an additional individual microresistor. Thus, in the device, the signal is first amplified by an avalanche process in a pixel, and then by a microtransistor (an "individual emitter - semiconductor region - semiconductor layer" structure) made on the surface of this pixel. The signal is taken from an external load resistance connected to the electrical circuit of an additional conductive bus. The overall signal gain is defined as $M_0$=$M_av$·$M_tr$, where $M_av$ is the gain of the avalanche process, $M_tr$ is the gain of the microtransistor.

The semiconductor avalanche detector is implemented as follows. On the surface of the semiconductor layer 1, for example, an n-type silicon layer with a resistivity of 2 Ohm cm, a dielectric layer 5 of silicon dioxide ($SiO_2$) with a thickness of 0.1 u is formed by thermal oxidation at a temperature of 1100 C. Windows of 40 u×40 u in size and with an interval of 10 um are opened on the surface of the oxide by a photolithographic method. Then, semiconductor regions 2 (pixels) of p-type conductivity are formed in these windows by ion doping with boron at a dose of 10 uC/$cm^2$ and an energy of 70 keV. After thermal dispersal of boron to a depth of 1.5 u, on a small surface (about 5 u × 5 u) of each pixel, an individual emitter is formed by ion doping with phosphorus at a dose of 150 uC/$cm^2$ and an energy of 100 keV. The distillation of phosphorus is carried out to a depth of 0.7u. Contact regions to pixels are formed by additional doping of a small area of semiconductor regions with boron ions at a dose of 50 uC/$cm^2$ and an energy of 70 keV. Microresistors with a surface resistance of about 20 ohms/$cm^2$ are made from amorphous silicon by vapour deposition. A common conductive bus and an additional bus are made of a two-layer metal (Ti + Al) by ion-plasma sputtering. Ohmic contact 9 to the semiconductor layer is formed by deposition of an aluminium layer on the free front surface of the semiconductor layer. Due to the low level of optical crosstalk and high speed, the proposed semiconductor avalanche detector can be widely used as detectors of light quanta and charged particles both in fundamental research (nuclear physics, high energy physics, etc.) and in applied fields (ecology, dosimetry), medical tomography, etc.)\cite{sadigov2022improvement}, \cite{sadygov2018photodetector}, \cite{sadygov2013technology}.

\section{Performance of photodetector.}

The need to improve the speed of micropixel avalanche photodetectors is caused by the possibility of their use in a number of nuclear physics detectors, new generation positron emission tomographs and in LIDAR systems that require time-of-flight modules capable of measuring the time of arrival of charged particles, gamma rays or photon beams with an accuracy no worse than than 100ps. For such modules, the duration (or slew rate) of the leading edge of the signal taken from the photodetector is important. The results of the study led to the development of a new design to improve the performance of micropixel avalanche photodetectors.

In order to improve the speed of the LIDAR detector, it was revealed that it was necessary to create a fundamentally new avalanche photodetector\cite{sadigov2016iterative}. The fact is that, being a semiconductor device, MAPD photodetectors have a high specific capacitance (from 10 to 50 pF/$mm^2$). This leads to a limitation of the instrument's performance with an increase in its sensitive area. The micropixel avalanche phototransistor (MAPT) solves this problem. MAPT contains a matrix of micropixels with quenching resistors separate for each micropixel, as known MAPD, and a matrix of microtransistors with separate ballast resistors. All micropixels are connected via quenching resistors to a common metal conductor, and all microtransistors are connected via ballast resistors to another metal conductor. That is, the device has two independent signal outputs: a “pixel output” connected to a micropixel circuit and a “transistor output” connected to a microtransistor circuit (see Fig. 2).

\begin{figure}[h]
    \centering
    \includegraphics[width=0.5\textwidth]{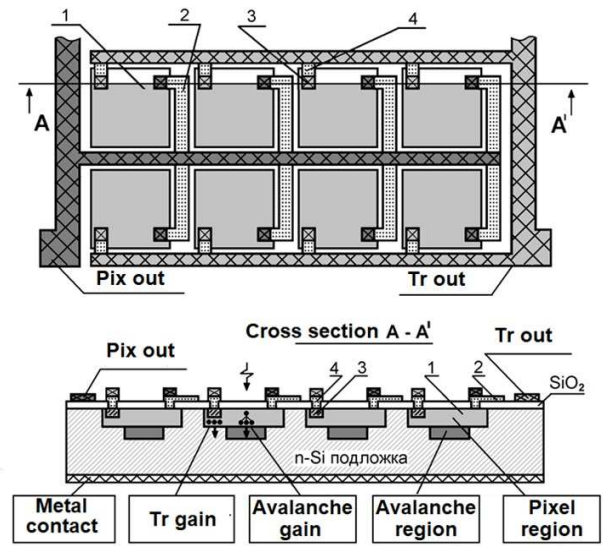}
    \caption{Design of a micropixel avalanche phototransistor fabricated on an n-type silicon substrate. 1-Micropixel, 2-Micropixel Resistor, 3-Microtransistor, 4-Microtransistor Resistor.}
    \label{fig1}
\end{figure}

The MAPT principle of operation is based on the specifics of the MAPD pixel operation in the Geiger mode (in the overvoltage mode)\cite{heydarovnew}. The avalanche process in the pixel causes a voltage drop there greater than 0.75V, which is enough to completely open the emitter-base (or pixel) junction of the microtransistor, and as a result of this, a current flows through the microtransistor's electrical circuit, limited only by its ballast resistor. The microtransistor closes when the potential of the base (pixel) relative to the emitter drops below 0.75V, due to the discharge of the micropixel capacitance. A microtransistor is formed directly over a small portion of the pixel area. The microtransistor has a size of 4u*4u which occupies about 3\% of the area of a 25u*25u micropixel. In this case, the specific capacitance of the microtransistor will be 80 times less than the corresponding capacitance of the pixel, which is equivalent to a decrease in the effective capacitance of the device, since the signal is taken from the microtransistor circuit (Figure 3 A). For example, if the overvoltage is 1V, the maximum potential change per pixel is 2V. It can be calculated that the photocurrent $J_p$ in the pixel circuit leads the current $J_tr$ in the microtransistor circuit, since the beginning and end of the latter is determined by the potential at the pixel (ie, the potential at the base). The microtransistor opens at the time t, when the avalanche process has time to discharge the pixel by $\vec{\Delta}$$U_dp$=$U_d$-$U_p$ (t)=0.75V. Thus, the photocurrent $J_p$ in the pixel circuit and the current $J_tr$. in the microtransistor circuit can be described by the expressions

\[J_p=\vec{\Delta}U_dp/R_q\] 
\[J_tr=(\vec{\Delta}U_dp-0.75)/R_tr\]

where $U_dp$=$U_d$-$U_p$ (t) is the potential change at the pixel, $U_d$ is the external voltage applied to the MAPT, $U_p$ (t) is the current potential at the pixel.
Methods for picking up a fast signal from MAPD (or SiPM) pixels using an independent electrode are known in the scientific literature. For example, special microcapacitors formed on a part of the pixel surface are used for this. All micro-capacitors are connected to an additional metal bus, which has an independent output for removing the fast signal (“fast out”). 
The closest commercial prototype of the developed photodetector is SiPM from SenSl company which formed from a large number (hundreds or thousand) trace elements (Figure 3 B). Each microcell is an avalanche photodiode with own quenching resistor and capacitive combined quick exit. These microcells are arranged in a densely packed array with all the same leads (all anodes) summed together. Thus, the array of microcells can be considered as a single photodiode sensor with three outputs: an anode, a cathode, and a fast output. The photodiode uses an additional electrode to output a signal, while the other two electrodes are used to ensure the device's operability, i.e. for food supply. The output provides a high reduction in both release time and rise time. This electrode is insulated by a dielectric from all other elements of the structure and has a weak capacitive coupling with each photosensitive pixel. Such a connection provides a fast partial injection of an electric charge when a pixel is activated by a photon hitting it. The capacitance of the third electrode in relation to other electrodes of the photodetector is very low and amounts to one tenth of the total capacitance. When creating this electrode, a transparent conducting oxide was used.

\begin{figure}[h]
    \centering
    \includegraphics[width=0.6\textwidth]{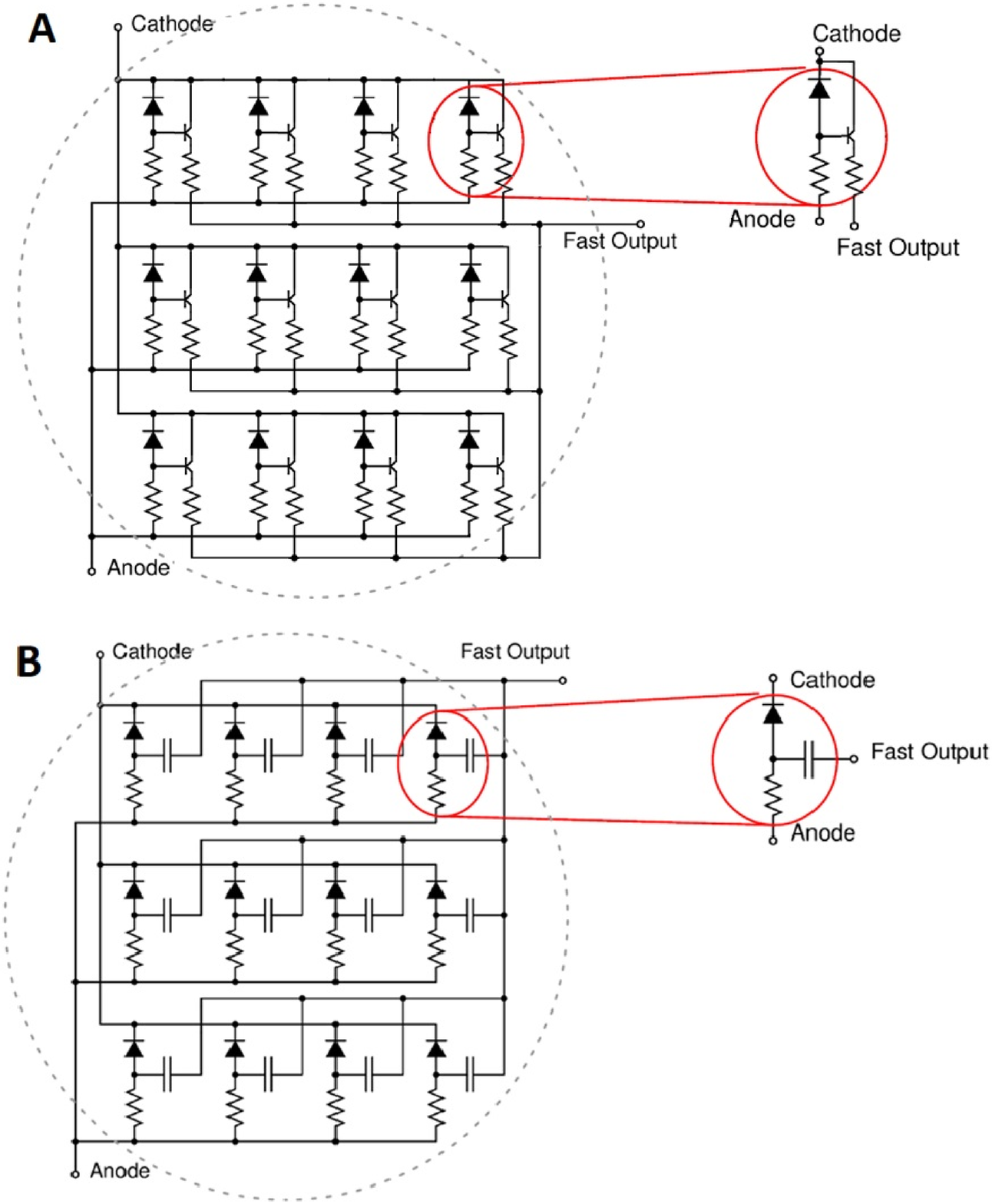}
    \caption{SiPM electrical equivalent circuits from MAPT (A) and SensL (B).}
    \label{fig1}
\end{figure}

The experimental data presented in Figure 4 show a significant advantage of micropixel avalanche phototransistor over SiPM from SensL company. The comparison of the two instruments was carried out as follows. Appropriate bias voltages were applied to the devices, at which the gain in both devices reached M=1.5*$10^6$. The red curve shows the photo signal taken from the transistor output of the photodiode developed by us with transistor amplification, the blue curve shows the photo signal taken from the capacitive output (or fast output) of the photodiode from SensL. Both signals were amplified by two identical amplifiers with a bandwidth of 80 MHz and a gain of 100, then fed to oscilloscope with a bandwidth of 200 MHz. In this case, the edges of the signals were determined by the bandwidth of the amplifier, but this does not impair the accuracy of the comparison. 
The amplitude of the signal taken from the fast output of the developed MAPT can be significantly increased by increasing the overvoltage and reducing the resistance of the ballast resistor $R_tr$. In this case, the growth rate (mV/ns) of the leading edge of the signal will increase significantly, which will improve the accuracy of timing in LIDAR systems.

\begin{figure}[h]
    \centering
    \includegraphics[width=0.4\textwidth]{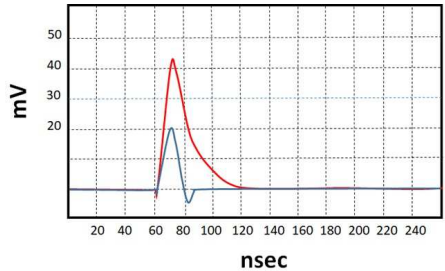}
    \caption{Oscillogram of photodiodes. The red curve is the photo signal from the MAPT, the blue curve is the photo signal of the photodiode from the SensL company}
    \label{fig1}
\end{figure}

The table bellow shows the results of comparing the parameters of the developed MAPT with its analog from the SensL company (instrument type: MicroPC-30035-SMT), in which a fast signal is taken from a chain of microcapacitors  located on the surface of micropixels. 

\begin{tabularx}{0.8\textwidth} { 
  | >{\raggedright\arraybackslash}X 
  | >{\centering\arraybackslash}X 
  | >{\raggedleft\arraybackslash}X | }
 \hline
Parameter & SenSl & MAPT \\
 \hline
Amplitude of signal  & 1.2mV  & 2.8mV  \\
  \hline
Gain  & 0.63*$10^5$  & 5.1*$10^5$  \\
   \hline
Rise rate of signal edge  & 2,4*$10^5$ V/sec.  & 5*$10^5$ V/sec.  \\
   \hline
Sensitive area  & 3mm*3mm & 3mm*3mm  \\
\hline
\end{tabularx}

It can be seen that the main parameters of the signal taken from the fast output of the MAPT significantly exceed the corresponding parameters of the SensL device.

\section{Results}
A new photodetector has been developed and experimentally implemented - a micropixel avalanche phototransistor, containing a matrix of photodiode pixels, in which each pixel is connected to an individual n-p-n microtransistor operating in the digital mode. It has been experimentally established that the MAPT is significantly superior to known analogues in the world in terms of speed and sensitivity. It is shown that the amplitude of the signal taken from the fast output of the MAPT can be significantly increased by increasing the overvoltage on the device and reducing the resistance of the microtransistor ballast resistor.

\section*{Acknowledgement}
This work was supported by the Scientific Foundation of SOCAR. 

\printbibliography

\end{document}